\documentclass[bibyear]{aa} 

\usepackage{graphicx}

\usepackage{txfonts}
\begin{document}

  \title{Rotationally and tidally distorted compact stars }
 { }
   \subtitle{A theoretical approach to the gravity-darkening exponents for white dwarfs}

\author{A. Claret \inst{1, 2}}
   \offprints{A. Claret, e-mail:claret@iaa.es. }
\institute{Instituto de Astrof\'{\i}sica de Andaluc\'{\i}a, CSIC, Apartado 3004,
            18080 Granada, Spain
            \and
            Dept. F\'{\i}sica Te\'{o}rica y del Cosmos, Universidad de Granada, 
            Campus de Fuentenueva s/n,  10871, Granada, Spain}
            \date{Received ; accepted; }

\abstract
{To the best of our knowledge, there are no specific calculations of  gravity-darkening 
exponents for white dwarfs in the literature. On the other hand, the number 
of known eclipsing binaries   whose components 
are tidally and/or rotationally distorted white dwarfs is increasing year on year.}
{Our main objective is to present the  first 
        theoretical approaches to the problem 
of the distribution of temperatures on the surfaces of compact stars distorted 
by rotation and/or tides in order to compare with relevant  observational data.}
{We
used two methods to calculate  the gravity-darkening exponents: (a) a variant of our numerical method based on the triangles 
strategy and (b) an analytical approach consisting in a generalisation 
of the von Zeipel theorem for hot  white dwarfs.  }
{ We find discrepancies between the gravity-darkening exponents calculated with 
our methods and the predictions of the von Zeipel theorem, particularly 
in the cases of cold white dwarfs; although the discrepancy also applies to higher 
effective temperatures under determined physical conditions.
We find physical connections between the gravity-darkening 
exponents calculated using our modified method of triangles strategy with 
the convective efficiency (defined here as the ratio of the convective to the 
total flux). A connection between the  entropy and the 
gravity-darkening coefficients is also found: variations of the former cause 
changes in the way the temperature is distributed on distorted stellar surfaces. 
 On the other hand, we have generalised the von Zeipel theorem  
for the case of hot white dwarfs. Such a generalisation allows us to 
predict that, under certain circumstances, the value of the gravity-darkening 
exponent may be smaller than 1.0, even in the case of high effective temperatures.
}
{ To constrain the gravity-darkening exponent values observationally it would be necessary 
to find and investigate eclipsing binaries constituted by white dwarfs showing tidal and/or rotational 
distortions that were double-lined and that were bright enough to obtain 
good radial-velocity semi-amplitudes for both components. It would be very interesting 
and useful if observer were to  focus their attention on this kind of system 
to check our theoretical results. 
}

   \titlerunning {  }
   \maketitle
%

\section{Introduction}

Gravity-darkening is  an important piece of the stellar structure and evolution 
that has been studied for almost a hundred years (von Zeipel 1924). Gravity-darkening 
exponents (GDE) are key tools for analysing light curves of eclipsing binaries or 
in isolated rotating stars through long-baseline optical and infrared interferometry.  
 Considering  stars  in strict radiative equilibrium 
(pseudo-barotrope), in 1924 von Zeipel showed that the variation of brightness 
over the surface is proportional to the effective gravity, that is,

\begin{equation}
        {F} = -{4 a c T^{3}\over{3 \kappa \rho}}{dT\over{d\psi}} {g^{\beta_{1}}}
,\end{equation}

\noindent
or equivalently 

\begin{equation}
        {{T_{\rm eff}}^4} \propto {g^{\beta_1}}, 
\end{equation}

\noindent
where  $\psi$ is the potential, $g$ the local gravity, T  the local 
temperature, $\kappa$  the opacity, $\rho$  the local density, $a$  
the radiation pressure constant, $c$ the velocity of light in vacuum, T$_{\rm eff}$  
the effective temperature, and $\beta_1$=1.0 is the GDE, which is  a bolometric quantity.
Although we are still  far from fully understanding the gravity-darkening phenomenon, 
in the last 22 years several 
theoretical papers related to the GDE have appeared in the literature that have 
shed some light on the scenario, such as Claret (1998, 2000, 2012), Espinosa 
Lara and Rieutord (2011, 2012), and McGill, Sigut, and Jones (2013). For a historical 
summary of the theoretical research on the GDE see Claret (1998, 2015). 
From now on we  also designate the GDE as $\beta_1$. 

Another  important and complementary ingredient in the analysis of 
systems with non-spherical configurations are the so-called gravity-darkening 
coefficients (GDCs).   As we can only observe
        determined band-limited stellar flux (not the bolometric), the
        introduction of these coefficients is necessary in order to model a distorted star. Such a coefficient can be written as (Claret\& Bloemen 2011):

\begin{eqnarray}
        y(\lambda, T_{\rm eff }, \log [A/H], \log g, V_{\xi}) =  \nonumber\\
        \left(\frac{d\ln T_{\rm eff }}{{d\ln g}}\right)
        \left(\frac{\partial{\ln I_o(\lambda)}}{{\partial{\ln T_{\rm eff }}}}\right)_{g}
        + \left(\frac{\partial{\ln I_o(\lambda)}}
        {\partial{\ln g}}\right)_{T_{\rm eff}},
\end{eqnarray}
\noindent
\noindent
where $\lambda$ is the wavelength,  $I_o(\lambda)$ the intensity at a given 
wavelength  at the centre of the stellar disc, and $V_{\xi}$ is the  microturbulent  velocity.  
We note that the expression $\left(\frac{d\ln T_{\rm eff }}{{d\ln g}}\right)$ can 
be written as $\beta_1/4$.
In order to progress in our investigation of stellar 
configurations distorted by rotation and/or tides, we recently studied 
the GDCs for the case of compact stars (Claret et al. 2020). In that study 
we computed GDC for DA, DB, and DBA white dwarf models, covering the 
transmission curves of the Sloan, UBVRI, Kepler, TESS, Gaia, and HiPERCAM 
photometric systems. These computations are available for log [H/He] = 
-10.0 (DB), -2.0 (DBA), and   He/H = 0 (DA). The covered range 
of log g was 5.0-9.5, while for the effective temperatures the respective 
range was 3750 K-100000 K.

From an observational point of view, discoveries of binary systems whose components
are tidally and/or rotationally distorted white dwarfs are ongoing (see e.g. 
Burdge et al. (2019a, 2019b, 2020), Kupfer et al. (2020)  and references therein).
However, as far as we know, there are no specific GDE calculations 
for white dwarf sequences or even for an individual model. 
In this short paper we present, for the first time, GDEs for three 
white dwarf cooling sequences. We also introduce some improvements to 
our methods for calculating the GDE:   
$\beta_1$ is computed as a function of the optical depth, that is, $\beta_1 = 
\beta_1(\tau)$. In addition, we introduce an extra  condition in our 
calculations, namely the relationship between the local gravities and the corresponding 
optical depths for a given  equipotential surface. 
In the following, sections we discuss these points in more detail.

The paper is organised as follows:  Sect. 2 is dedicated to describing our  methods  and some 
aspects of the distorted stellar configurations. In Sect. 3 we discuss both observational and 
theoretical evidence for deviations from von Zeipel's  theorem (1924) and 
present our  results and conclusions.

\section{ The numerical method}

Our numerical method is based on the triangles strategy 
introduced by Kippenhahn et al. (1967). A complete description of our method can be 
found in Claret (1998), but for the sake of clarity, we summarise it below.  To save 
computing time, Kippenhahn et al. (1967) introduced an ingenious method: when an 
evolutionary sequence is being calculated, if the external boundary conditions are 
unchanged at the fitting point (envelope-interior), the outer layer integrations 
must be the same as the previous ones. Three envelopes corresponding to three points 
in the  Hertzsprung-Russell (HR) diagram around the current values of luminosity and effective temperature are computed. 
As the model evolves, its properties are checked to see if they remain within this
triangle. If so, the boundary conditions are unchanged. It is important to highlight that 
this strategy is valid only if the triangle is sufficiently small. This warning is 
particularly valid for convective envelopes.  To simulate a distorted star, we are 
interested in several triangles showing different effective temperature distributions 
over the surface. Therefore, we increase the number of triangles in the HR diagram, that is, 
we compute several envelopes with different temperature distributions 
but imposing the same physical conditions at a given interior point. Figure 2 
in Claret (2000) shows a simplified scheme of our method where only three points 
are shown in the HR diagram. Once the triangles have been computed  for 
each point of the evolutionary track we can derive $\beta_1$ by differentiating 
the neighbouring envelopes. Such a procedure is performed for the next evolutionary 
track point and so on. 

Our method has some advantages: (a) it can be applied to  convective and/or 
radiative envelopes; (b) one can investigate the influence of the optical depth 
in the GDE by changing the fitting point to impose the boundary conditions, 
without loss of generality; (c) the GDE can be computed as a function of initial mass, 
chemical composition, evolutionary stage (in the present paper up to the white 
dwarf phase), and other ingredients of the input physics, and (d)  more realistic atmosphere 
models can  easily be incorporated as external boundary conditions, 
as done in Claret (2012). 

We introduce  an extra condition  in our procedure:   the 
relationship between local gravity and optical depth over a given equipotential surface. 
In reality, solving the hydrostatic equilibrium equation for two points on an equipotential 
characterised by [$g(\mu), \tau(\mu,\psi)$] , 
we  obtain $g(\mu)\tau(\mu, \psi) = g(\mu_o) \tau (\mu_o, \psi)$.  In this relationship, 
 $g$ is the local gravity, $\psi$ is the total potential (rotational one included), 
$\tau$ is the optical depth, and $\mu$ is given by $\cos(\phi)$ where $\phi$ 
is the angle between the radiation field and the $z$ axis.  The  subscript $o$ 
indicates the reference point, for example the pole. 
To guarantee that the triangle technique represents an equipotential, we have taken 
as reference the absolute dimensions for each point of the evolutionary models, 
that is, [$g(\mu_o), \tau(\mu_o,\psi) $]. The equipotentials are then configurated introducing 
 additional triangles from this point.

 As outlined in Sect. 1, for stars in strict radiative equilibrium, von Zeipel's theorem predicts an exponent $\beta $ = 1.0. However, if we inspect a typical HR diagram in the log g  $\times$ T$_{\rm eff}$ plane  
for stars with  different initial masses, for example, 1.0 M$_{\odot}$ (convection 
predominates in the envelope) and 10.0 M$_{\odot}$  (envelope predominantly in radiative 
equilibrium), it can be verified that both models have different average slopes. 
If we make a simple analogy between these two slopes and the GDE given by Eq. 2, the 
respective $\beta_1$ would be different, with the one corresponding to the star with 
1.0 M$_{\odot}$ being smaller than that of the model with 10.0 M$_{\odot}$; see Fig. 3 by Claret (1998) for a graphic example. This figure seems to  indicate that stars with convective 
envelopes do not strictly obey von Zeipel's theorem.  Additional and more elaborate evidence has also been found to support these statements, 
such as that outlined in  Claret (2012, 2015) for example. In these latter studies,  significant deviations were 
found when the GDEs are computed for the upper layers of a distorted star. 

On the other hand, Kopal (1959) derived the following  equation for the  stellar distortions:

\begin{eqnarray}{{g-g_o}\over{g_0}}={\sum_j}\left(1-{5\over{\Delta_j}}\right)
\left({r\over{a_1}}-1\right)  
,\end{eqnarray}

\noindent
where g$_o$ is the reference local gravity,   $\Delta_{j}$ = 1 + 2 k$_j$  
, and k$_j$ is the apsidal motion constant 
of order $j$. Therefore, the radius of an equipotential $r$ (order 2) 
can be written as

\begin{equation}
r = a_1\left(1 - f_2 P_2(\theta, \phi)\right),   
\end{equation}

\noindent
with  

\begin{eqnarray}f_2 = {5\omega^2a_1^3\over{3GM_{\psi}(2+\eta_2)}},   
\end{eqnarray}

\noindent
where $\omega$ is the angular velocity, P$_2(\theta, \phi)$ 
is the second surface harmonic, $a_1$ is the mean radius of the level surface, 
$\eta_2$ is the logarithmic derivative of the amplitudes of the surface distortions defined 
through Radau's differential equation, and $M_{\psi}$ is the mass enclosed by an 
equipotential. Claret (2000) has shown that there is a close connection 
between the GDE and the shape of the distorted stellar configuration,
its internal structure, and the details of the rotation law. 
In fact, for stellar masses around 1.5 M$_{\odot}$ there is a change in the 
predominant source of thermonuclear energy from the proton--proton chain to the CNO 
cycle. This change causes  a readjustment of the mass 
concentration through the parameter $\eta_2$. 
 We reiterate the fact that $\eta_2$ is connected to k$_2$   
through a simple equation: $k_2=(3-\eta_2(R))/(2(2+\eta_2(R)))$, where 
$\eta_2(R)$ is evaluated at the stellar surface.  
This readjustment will affect 
how a star reacts under distortions and  will consequently effect the parameter  $\beta_1$  (see Eq. 3   and also  Fig. 1 by Claret (2000) where log k$_2$ and $\beta_1$  
are shown for ZAMS models with masses varying from 0.075 to 40.0 M$_{\odot}$). 
In addition,  convection also  begins to contribute significantly 
to the total flux in this range of effective temperatures.  
For masses greater than 1.5 M$_{\odot}$ the mass concentration decreases almost 
linearly with the stellar mass.

\section {Discussion of the results and final remarks}

Some years ago, Claret (2016, see the corresponding Eqs. 8 and 9) adopted  
an expression for the convective efficiency (ratio of the convective to the 
total flux, denoted by the symbol $f$) to investigate the GDEs. 
Here  we  generalise that result for a range of opacities 
(through parametrised formulae) as 
indicated by Eq. 7:

\begin{eqnarray}
{f} \approx A \gamma \left({r\over R}\right)^2  \left({3 \Gamma_1\over{5 \mu_1 \beta}}\right)^{1/2} 
\left({2 c_P \mu_1 \beta\over{5}}\right) T^{1/2} 
\left[ g\over{T_{\rm eff}^{{(4n+4+\left|n+s\right|)}}} \right].
\end{eqnarray}

\noindent
In the above equation, $A$ is a constant, $R$ the star radius, $r$ the 
radial coordinate, $\mu_1$ 
 the mean molecular weight,  $\beta$  the ratio of gas to total pressure,  
$c_P$  the specific heat at constant 
pressure, and $\Gamma_1$ is the 
adiabatic exponent.  The parameter $\gamma$ is given by

\begin{eqnarray}
{\gamma} = {1\over 2} {\overline{v}\over{v_s}} {\Delta T\over T}, 
\end{eqnarray}

\noindent
with $\overline{v}$ being the mean convective velocity,   
$\Delta$T  the excess of temperature of a rising element over the 
mean temperature of the surrounding environment, and $v_s$ is the  velocity  of sound. 
Here, we assume that for the derivation of Eq. 7 the opacity can be written as  
$\kappa \approx \kappa'  P^{n} T^{-s}$ or $\kappa\approx\kappa''\rho^{n} T^{-n-s}$, 
where   $\kappa'$ and  $\kappa''$  are constants, and  $n >$ 0 and $s <$ 0, 
assuming a perfect gas equation. 
 Although we will not use this expression directly because it is only 
an approximation, it is useful in order to analyse the correlation between $f$ ---through 
the role of the  opacities--- and $\beta_1$, as we see in the following paragraphs.

On the other hand, the evolutionary track from the pre-main sequence (PMS) up to the white dwarf stage 
was computed using the MESA module (Paxton et al. 2011, 2013 and 2015), version 7385. 
The basic input physics of the MESA code is given in the
above references. 
Another set of DA- and DB-type white dwarf models was computed with 
LPCODE (Althaus et al. 2001a, 2001b). The models were followed from  the 
zero age main sequence (ZAMS) up to the white dwarf stage (Renedo et al. 2010). 
This code considers modern input physics such as a detailed network for 
thermonuclear reactions, OPAL radiative opacities, full-spectrum turbulence 
theory of convection, detailed equations of state, and
neutrino emission rates.

We  first discuss the 
cases of `pure' DA- and DB-type white dwarfs, without considering the previous 
evolutionary phases. We  analyse models with 1.0 M$_{\odot}$ although the results are 
similar if we consider models with  different initial masses. Figure 1 illustrates the resulting GDE 
computed at optical depth $\tau$= 100.0 (thick continuous line) using the modified triangles strategy method. The log g  during the  evolution 
of this cooling sequence vary approximately between 8.3 and 8.7. 
As expected, for high effective temperatures where radiative transfer 
predominates and at large optical depth, the results are compatible with the equation of diffusion, that is, with those resulting from the  
von Zeipel (1924) approach. However, as the model evolves, the influence of convection 
begins to appear which translates into a drop of $\beta_1$. 
The drop-off threshold is located at logT$_{\rm eff}$ $\approx$ 4.12 and 
 we have found large deviations from 
the von Zeipel theorem  for  effective temperatures smaller than 10000K.  
We note that this transition temperature is slightly  different from that typical of main sequence 
stars (logT$_{\rm eff}$ $\approx$ 3.90). This is due to differences in chemical 
compositions and degree of compactness (log g).
The dashed (log g = 8.5) and thin continuous (log g=8.0)  lines are useful 
to establish a more direct connection between 
the contribution of the convection  to the flux and $\beta_1$.   Such a  contribution 
can be set in terms of the ratio F$_c(\tau)$/F$_r(\tau)$  
where F$_c(\tau)$ is the convective flux and F$_r(\tau)$ is the radiative one for a 
given optical depth (these data were kindly provided by  E. Cukanovaite 2020; 
see also Cukanovaite et al. 2019). 
The F$_c(\tau)$/F$_r(\tau)$ ratio is slightly different from that given by Eq. 7 
but shows the same appearance, taking into account the different scales: the F$_c(\tau)$/F$_T(\tau)$ ratio  
varies from 0 to 1.0, F$_T(\tau)$ being the total flux. 

We can use Eq. 7 
 to help us visualise the behaviour of $\beta_1$ and its relation to opacity. 
The connection between $\beta_1$ and the convective contribution to the  
 flux is evident; it is  especially clear near the maxima and minima.    
The behaviour of the GDE shown in Fig. 1 (also that of the following figures) 
is the result of the combination of the different sources of opacity  
but we can gain some insight into its influence on the GDE by adopting some opacity laws in their parameterised form.
Firstly, we  analyse  the effects of  opacity in 
the colder models. For this range of effective temperatures 
one of the predominant sources of opacity is the negative hydrogen ion whose dependence 
on $\rho$ and T is given by  $\kappa \approx \kappa_1  \rho^{1/2} T^{7.7}$, where 
$\kappa_1$ is a constant. We  note the high dependence of  negative hydrogen ion  opacity 
with temperature.  
The expression 
$ g\over{T_{\rm eff}^{{(4n+4+\left|n+s\right|)}}}$ 
in Eq.  7 drives the behaviour of the gravity-darkening exponent, and,  
introducing the relevant values of $n$ and $s$ 
 for such models at a given convective efficiency,  we get $\beta_1\approx$ 0.30. 
 In addition,  the opacities $ff$, $bf,$ and 
 $bb,$ because of electronic transitions, are given by Kramers law:  
 $\kappa \approx \kappa_2  \rho T^{-7/2}$, where 
 $\kappa_2$ is a constant. The approximate corresponding  value of $\beta_1$ 
 in this case is $\approx$ 0.35. 
Considering that Eq. 7 is just a rough approximation, these results are surprising because they  
predict  deviations from the von Zeipel's theorem and, in addition, they are in 
reasonable agreement with the values of $\beta_1$  found in the literature for  
late-type stars (semi-empirical and theoretical ones; see below). On the other hand, one of the 
main sources of continuum opacity in hot star  atmospheres is the so-called 
Thomson scattering. As known, this opacity is `grey' because there is no 
dependence on temperature and density and it can be written as follows for the case of complete ionisation:  
$\kappa \approx 0.2 (1.0+X)$, where $X$ is the hydrogen content. Using the same 
treatment as for the negative hydrogen ion case and Kramers law, we obtain 
$\beta_1 \approx 1.00$, which is in good agreement with the predictions of the 
von Zeipel's theorem for hot stars. Again it is gratifying that Eq. 7  is 
capable of predicting 
the typical values of $\beta_1$, even considering its limitations.

In addition to the analysis of the influence of convective flux on $\beta_1$, 
there are other ways to correlate $\beta_1$ with some physical magnitudes. 
Considering the additive properties of the specific entropy in the nonrelativistic 
case we have

\begin{eqnarray}{s} = B+ {N_ok\over{\mu_i}}ln{T^{3/2}\over{\rho}} + 
        {{N_ok\over{\mu_e}}} \left[ {5\over{3}} {F_{3/2}(\alpha)\over{F_{1/2}(\alpha)}} 
        + \alpha\right]   
        + {4a\over{3}} {T^3\over{\rho}}    
,\end{eqnarray}

\noindent
where the symbols have the following meaning: B is a constant, $\alpha$ the 
degeneracy parameter,  N$_o$ is Avogadro's number, $k$ the Boltzmann constant, 
$\mu_i$ the mean molecular weight per ion, and   $\mu_e$ is the mean molecular 
weight per electron.  The functions $F_{1/2}(\alpha)$ and $F_{3/2}(\alpha)$ 
are auxiliary functions that can be written as 

\begin{eqnarray}
F_{1/2}(\alpha)= \int_{0}^{\infty} {u^{1/2}du\over{e^{\alpha+u} +1}},
\end{eqnarray}

\noindent
and

\begin{eqnarray}
        F_{3/2}(\alpha)= \int_{0}^{\infty} {u^{3/2}du\over{e^{\alpha+u} +1}},
\end{eqnarray}

\noindent
where $u=p^2/(2 m kT)$, with $p$ being the particle momentum and $m$ its mass. 
The three components in Eq. 9  above can be easily identified: the second term 
is the entropy due to ions, the third is connected to electrons, and the 
fourth to radiation. We can see in Fig. 2 how the GDE is related to 
 the  entropy for the same models and conditions 
shown in Fig. 1. As mentioned,  the convection onset is in 
log T$_{\rm eff}$ $\approx$ 4.12 (log T$_{\rm eff~onset}$) 
where a sudden variation of the entropy is indicated by a vertical arrow. This 
effective temperature  is in agreement with those   by Tremblay (2020).  
There are  three other inflexion points (also marked with vertical arrows). 
These four points shown in Fig. 2 indicate that the  entropy  varies with the 
effective temperature (also with log g), which in turn drives the  behaviour of the GDE.  
  The changes in $\beta_1$ with entropy are not surprising. 
Indeed, the differential of the entropy is given by  $dS=c_p (\nabla-\nabla_{adia})$,  
where $\nabla=dlnT/dlnP$ and $\nabla_{adia}=(dlnT/dnlP)_{adia}$. On the other hand, 
it can be shown that, alternatively, the  convective flux F$_c(\tau) \propto(\nabla-\nabla_{adia})^x$, 
where x = 3 (small convective efficiency) or x = 3/2 for large convective efficiency. Thus, under determined physical conditions, the entropy can also be used as a convective stability criterion.

Another point to note is that the  entropy does not significantly depend on the 
optical depth for effective temperatures $\leq$ T$_{\rm eff~onset}$. Indeed, the curves 
for $\tau $ = 100.0 and 500.0 coincide for T$_{\rm eff}$ $\leq$ T$_{\rm eff~onset}$. 
This implies that the behaviour of the resulting GDEs should not vary significantly, 
at least  within the range of optical depths, log g, and effective temperatures  explored here. We note that, for hot models, because
the  entropy 
hardly varies with T$_{\rm eff}$, the  corresponding GDE values are almost 
constant and equal to 1.0, which restores von Zeipel's  theorem. 
 These results are more general than  those obtained decades ago, whose GDE 
 value was constant and  approximately equal to 0.32 for envelopes in 
 convective equilibrium.

Figure 3  shows a comparison between the GDE for DB and DA models  at  
$\tau$ = 100.0. The profiles of $\beta_1$ are  very similar, with the exception 
of the transition zone where there is a  shift due to the difference in the  
chemical composition of the models and its influence on pressure and 
temperature.

An interesting comparison that can be made is related  to the masses of 
white dwarfs. In Fig. 4 we show the evolution of the GDE for DA models 
with masses 1 M$_{\odot}$  (continuous line) and 0.52 M$_{\odot}$ (dashed line). The 
calculations were also performed for $\tau$ = 100.0. Because of
the dependence of the onset of convection on the local gravity, the 
GDEs for the model with 0.52 M$_{\odot}$ are shifted towards lower temperatures,  
in reasonable agreement with other  studies (Cunningham et al. 2019, 
Tremblay (2020)).

As  indicated, another set of evolutionary models was generated with 
the MESA module with an initial mass of 2.0 M$_{\odot}$, X = 0.703, and Y = 0.277. 
For stars with convective envelopes, we employed the 
standard mixing-length formalism (B\"ohm-Vitense  1958) with $\alpha_{MLT}$ = 1.80. 
The opacity tables adopted are those given by Grevesse and Sauval (1998). 
 The models were followed from the PMS up to the cooling white dwarf stage. Figure 5  
shows the complete evolutionary track in the  HR diagram. The final mass in 
the cooling stage is 0.56 M$_{\odot}$. The results concerning gravity-darkening  
are shown in Fig. 6 where 
we add the profile of $\beta_1$ shown in Fig. 1 for comparison. 
Within our present level of approximation, the differences are small,  being more 
appreciable only in the interval around logT$_{\rm eff}$ $\approx$ 3.90.

As we can see in Figures 1-4 and 6, there are clear indications of deviations 
from the approach by von Zeipel.
Deviations from von Zeipel's theorem were previously 
found using suitable evolutionary models in stars evolving towards the branch 
of red giants and/or in   low-mass main sequence stars 
(low effective temperatures)  where the flux is predominantly convective 
in their envelopes (Claret 1998, 2000). 
Additional evidence for deviations from von Zeipel's theorem comes from 3D simulations 
of cold stars (Ludwig, Freytag\&Steffen 1999). 
A more complete historical review about this subject can be found in above  references. 

On the other hand, there is an analytical approach by which we can explore the behaviour 
of GDEs in compact stars whose outermost layers are in radiative equilibrium. 
Such an analysis was carried out in Claret (2012) 
for  main sequence stars. Here we adapt the physical conditions 
for the case of white dwarfs. Such an equation can be written 
as (see Appendix A)

\begin{eqnarray}
\beta_1 \approx  1 +  \left[{{2+N}-\chi(4+N)\over{2+N}} - 
{{2+N+\chi(-4+8\alpha_1-5N)}\over{2+N+\chi(\kappa_{\rho}-\kappa_T)(N-2\alpha_1)}}\right] 
\nonumber\\{\tau_p\over{\tau_e}}.
\end{eqnarray}

Equation 12 opens up some possibilities on the investigation of  
 the distribution of rotational velocities through the 
parameter $\alpha_1$ and on geometry through $N$. We note that 
this equation is valid only for hot  stars. 
We  note that the 
ratio $\tau_p/\tau_e \leq1.0$. For example, we would restore 
the predictions of von Zeipel's theorem  for the cases 
$\alpha_1= N/2$  or for uniform rotation 
and no $\theta$ dependence.  
An interesting feature of Eq. 12 can be explored and is 
 related to $\kappa_{\rho}$ and $\kappa_T$.  
Depending on the behaviour of these two variables, we could have values 
of $\beta_1$ significantly smaller than 1.0.\footnote{We have also found significant deviations from von 
Zeipel’s theorem using our modified numerical method,  at the upper layers of hot  white dwarfs.} 
For example, for $\alpha$=-1.0, N = 1, $\chi$ = 0.2, $\tau_p$/$\tau_e$= 0.71, 
and  $\beta_1$ = 0.80, and the resulting condition would be

\begin{eqnarray}
-6.0 \lessapprox \left[\left({{\partial ln\kappa}\over{\partial ln \rho}}\right)_{T} - 
  \left({{\partial ln\kappa}\over{\partial ln T}}\right)_{\rho}\right]  
  \lessapprox -5.0.
\end{eqnarray}

\noindent
Such a condition is approximately fulfilled for envelopes with effective 
temperatures in the interval 6000 K $\lessapprox$   T$_{\rm eff}$ $\lessapprox$ 13000 K. There are no semi-empirical  data yet for white dwarfs in this effective temperature range for comparison, but it is interesting to note that values 
of $\beta_1$ smaller than 1.0 were detected using long-baseline 
optical/infrared interferometry in isolated fast rotators. 
Some of these systems  show effective temperatures  within the 
mentioned range. 
A summary of their properties can be found in Table 1 of Claret (2016).

Values of the GDE smaller than 1.0 were also observationally  detected in main sequence and/or 
subgiants stars in binary systems.  
However,  a direct connection between the results from long-baseline 
 interferometry and from eclipsing binaries (mostly main sequence stars) and those 
provided by  Eq. 12 (compact ones)  is not straightforward.  However,  it can give us some clue for future research. 
A turning point concerning semi-empirical $\beta_1$ was the pioneering paper by Rafert\&Twigg (1980)  using eclipsing  
binaries. Such research was followed 
by others, such as Pantazis and Niarchos (1998), Niarchos (2000), and Djurasevic et al. 
(2003, 2006),  for example,  who explored a wide range of effective temperatures. 
The semi-empirical GDEs  
derived by these latter authors for hot stars are scattered around the classical 
von Zeipel value. Some of these systems show $\beta_1$ as low as 0.60.  Another 
important result of these investigations is that the derived values for systems 
with cooler components also contradict the predictions of von Zeipel's theorem. 
These semi-empirical GDE values are quite significant but do not yet 
constitute a critical test of the theory of temperature distribution on a 
distorted stellar surface. However, despite the fact that these 
observations are not conclusive, they seem to 
indicate a transition zone for the GDE. That zone approximately  coincides  
with the change in the prevalence of the process of energy transport in the envelopes, 
that is, radiative to convective, as indicated by the  Fig. 3 in Claret (2003). Such a transition zone is also  predicted in the present 
paper for compact stars.

Finally, it would be very interesting and useful 
if observers were to focus their attention on close binary systems constituted by 
white dwarfs distorted by rotation and tides, so that  the validity of the 
preliminary calculations presented here can be verified. 
To constrain the GDE values observationally, it would be necessary 
to investigate eclipsing binary white dwarfs that are double-lined and bright enough to obtain good radial-velocity  semi-amplitudes for both 
components. We hope that such systems will be found in the not too distant future.

\begin{figure}
\includegraphics[height=8.cm,width=6cm,angle=-90]{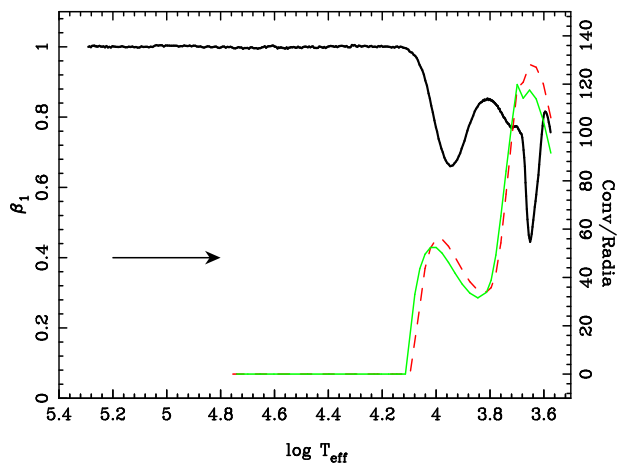}
\caption{DA-type white dwarf models (1.0 M$_{\odot}$). The thick solid line 
        represents the GDE as a function of effective temperature.   The continuous 
        thin line indicates the ratio F$_c(\tau)$/F$_r(\tau)$ for log g = 8.5  
and the dashed one denotes the same but for log g = 8.0. All  calculations 
were performed at  $\tau$ = 100.0. We note that the T$_{\rm eff}$  range   
of the atmosphere models  by Cukanovaite (2020) 
is limited to  Teff $\leq$ 60000 K. The arrow indicates the direction of time  
evolution.}
\end{figure}

\begin{figure}
        \includegraphics[height=8.cm,width=6cm,angle=-90]{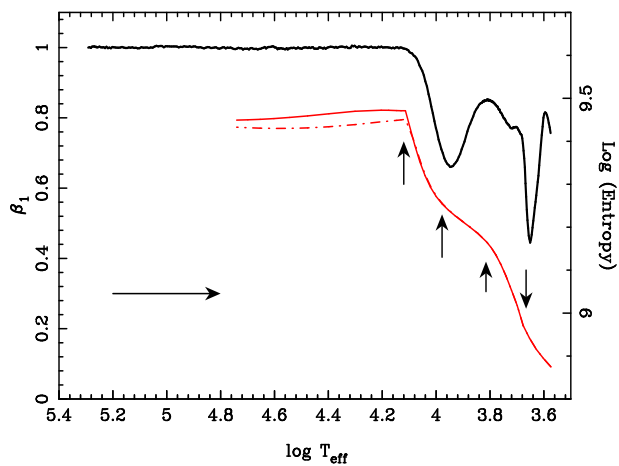}
        \caption{Entropy for the same models and conditions 
shown in Fig. 1 (red lines). The vertical arrows indicate the points where there 
is a marked  variation of the entropy with effective temperature. 
The solid line represents the entropy for $\tau$ = 100 while the 
dashed-dotted line denotes $\tau $ = 500;  both for log g = 8.5. 
 The horizontal arrow indicates 
the direction of time evolution.}
\end{figure}

\begin{figure}
\includegraphics[height=8.cm,width=6cm,angle=-90]{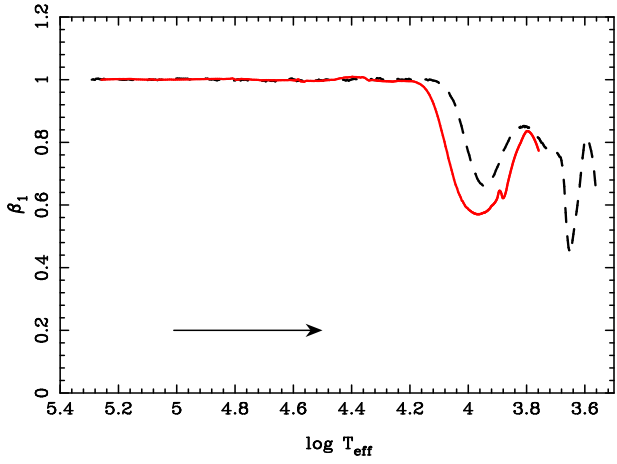}
\caption{Comparison between the GDE for a DB model (solid line, 1.0 M$_{\odot}$) 
and DA (dashed line, 0.5 M$_{\odot}$). All calculations were performed 
at $\tau$ = 100.0. The horizontal arrow indicates the direction of  time evolution.}
\end{figure}

\begin{figure}
\includegraphics[height=8.cm,width=6cm,angle=-90]{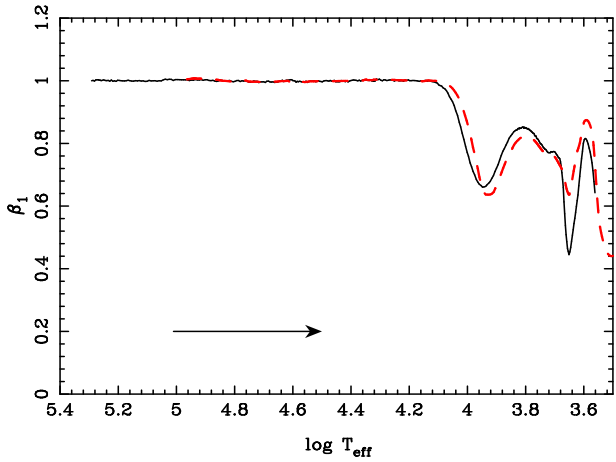}
\caption{Effect of log g on the onset of convection for DA models. 
The continuous line represents a sequence of DA model (1.0 M$_{\odot}$) 
while the dashed one also denotes  DA models but with 0.52 M$_{\odot}$. 
The horizontal arrow indicates the direction of time evolution.}
\end{figure}

\begin{figure}
\includegraphics[height=8.cm,width=6cm,angle=-90]{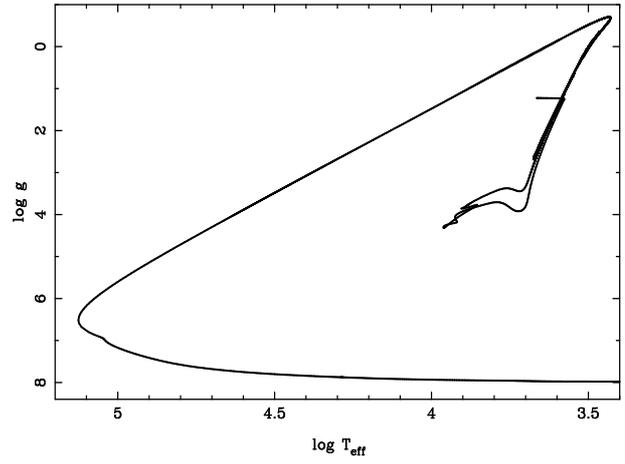}
\caption{HR diagram for an initial mass of 2.0 M$_{\odot}$ from 
the PMS to cooling white dwarf stage. Initial chemical composition 
X = 0.703, Y = 0.277, $\alpha_{MLT}$ = 1.80.}
\end{figure}

\begin{figure}
\includegraphics[height=8.cm,width=6cm,angle=-90]{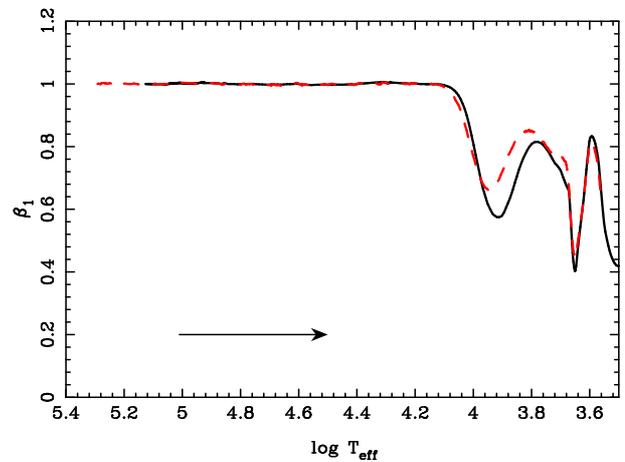}
\caption{GDE for  the white dwarf cooling sequence for the models shown 
        Fig. 5 (continuous line). 
The DA model  shown in Fig. 1 is represented by a dashed line. 
The horizontal arrow indicates the direction of time evolution. }
\end{figure}

\begin{acknowledgements} 
I would like to thank E. Cukanovaite for providing models of the 
structure of white dwarfs atmospheres  and an anonymous referee for his/her helpful suggestions. 
 The Spanish MEC (ESP2017-87676-C5-2-R,  PID2019-107061GB-C64, and  
PID2019-109522GB-C52) is gratefully acknowledged for its 
support during the development of this work. A.C. also 
acknowledges financial support from the State Agency for 
Research of the Spanish MCIU through the “Center of 
Excellence Severo Ochoa” award for the Instituto de 
Astrofísica de Andalucía (SEV-2017-0709). This research has made 
use of the SIMBAD database, operated at the CDS, Strasbourg, 
France, and of NASA's Astrophysics Data System Abstract Service.
\end{acknowledgements}

{}

\begin{appendix}
\section{Brief description of the derivation of Equation 12}

For completeness, we give here a brief summary of the derivation of Eq. 
12 which is a generalisation of von Zeipel’s theorem. For more details, see  
Kippenhahn (1977) and Claret (2012). 

Let us assume that the angular velocity has the following general form

\begin{equation}
{\omega^2} = {h(r) \sin^N\theta},   
\end{equation}

\noindent
where  $h(r)$ is a generic function and  $\theta$ is   the colatitude angle. In our approach  we assume that the effect of the centrifugal force on the stellar structure is small. 
The total potential $\psi$ can be written as  

\begin{equation}
\psi = {- G M\over{r}} - {r^2 \omega^2 \sin^2\theta\over{N + 2}}.  
\end{equation}

The integrating factor was chosen  in such a way that $\Lambda \vec{g}$ can be 
derived from a potential.  As a consequence, it can be shown 
that  the pressure is constant on  equipotential surfaces.
If the effects of rotation are small, 
we  obtain for the integrating factor:

\begin{equation}
{\Lambda} = 1 + {1\over{N+2}} \chi (N - 2\alpha_1) \sin^2\theta.  
\end{equation}

\noindent
In the above equation we define  $\alpha_1=
{\partial ln \omega\over{\partial ln r}}$ and $\chi = {r^3 \omega^2}/(G M_{\psi}$).   

On the other hand, the equation of the radiative transport can be written as 

\begin{equation}
{\vec{F} } =  {4ac\over{3 \kappa \rho}} T^3  {\vec{\nabla}}T.
\end{equation}

\noindent
Assuming that the density and temperature  are functions of the 
structural form of the potential we have ${\rho} = d(\Psi)\Lambda$ 
and $T = {t(\Psi)/{\Lambda}}$ and

\begin{equation}
{\vec{F} } =  {4ac t^3\over{3 \kappa d}} \left( {dt\over{d\Psi}} 
{\vec{g}\over{\Lambda^4}} + 
{t\over{\Lambda^6}} {\vec{\nabla}}\Lambda \right).
\end{equation}

As an extra condition we adopt the relationship 
$g(\mu_e)\tau(\mu_e, \psi) = g(\mu_p) \tau (\mu_p, \psi)$, 
where the subscripts $e$ and $p$ denote values at the equator and at the poles. 
 Calculating the values of  local gravity, 
of $\Lambda$ and of  Eq. A.5 at the pole and at the equator we finally 
obtain the GDE  through the logarithmic 
derivatives of the flux and effective gravity 

\begin{equation}
\beta_1  \approx {{\left({F_e-F_p}\over{F_p}\right)}\over
        \left({{g_e-g_p}\over{g_p}}\right)}. 
\end{equation}

After some algebra and incorporating the effects of variable opacity we finally get

\begin{eqnarray}
\beta_1 \approx 1 +\left[{{2+N}-\chi(4+N)\over{2+N}}-
{{2+N+\chi(-4+8\alpha_1-5N)}\over{2+N+\chi(\kappa_{\rho}-\kappa_T)(N-2\alpha_1)}}\right]  \nonumber\\{\tau_p\over{\tau_e}} 
,\end{eqnarray}

\noindent
where ${\kappa_{\rho}} \equiv \left({{\partial ln\kappa}\over{\partial ln \rho}}\right)_{T}$ 
and ${\kappa_{T}} \equiv \left({{\partial ln\kappa}\over{\partial ln T}}\right)_{\rho}$.

\end{appendix}

\end{document}